\documentclass[11pt,a4paper]{scrartcl}
  \usepackage[greek,pdf]{preamble}
  \usepackage{abstract}
  \title{\vspace{-2cm}Engines of Parsimony: Part~III}
  \subtitle{Performance Trade-offs for Reversible Computers Sharing Resources}
  \usepackage{extras}

\endofdump

  \def\largertitlepage{\thispagestyle{empty}\enlargethispage{2\baselineskip}}

\begin{document}

\maketitle
\largertitlepage

\begin{figure}[h!]
  \vspace{-1.5cm}
  \centering
  \includegraphics[width=.9\textwidth]{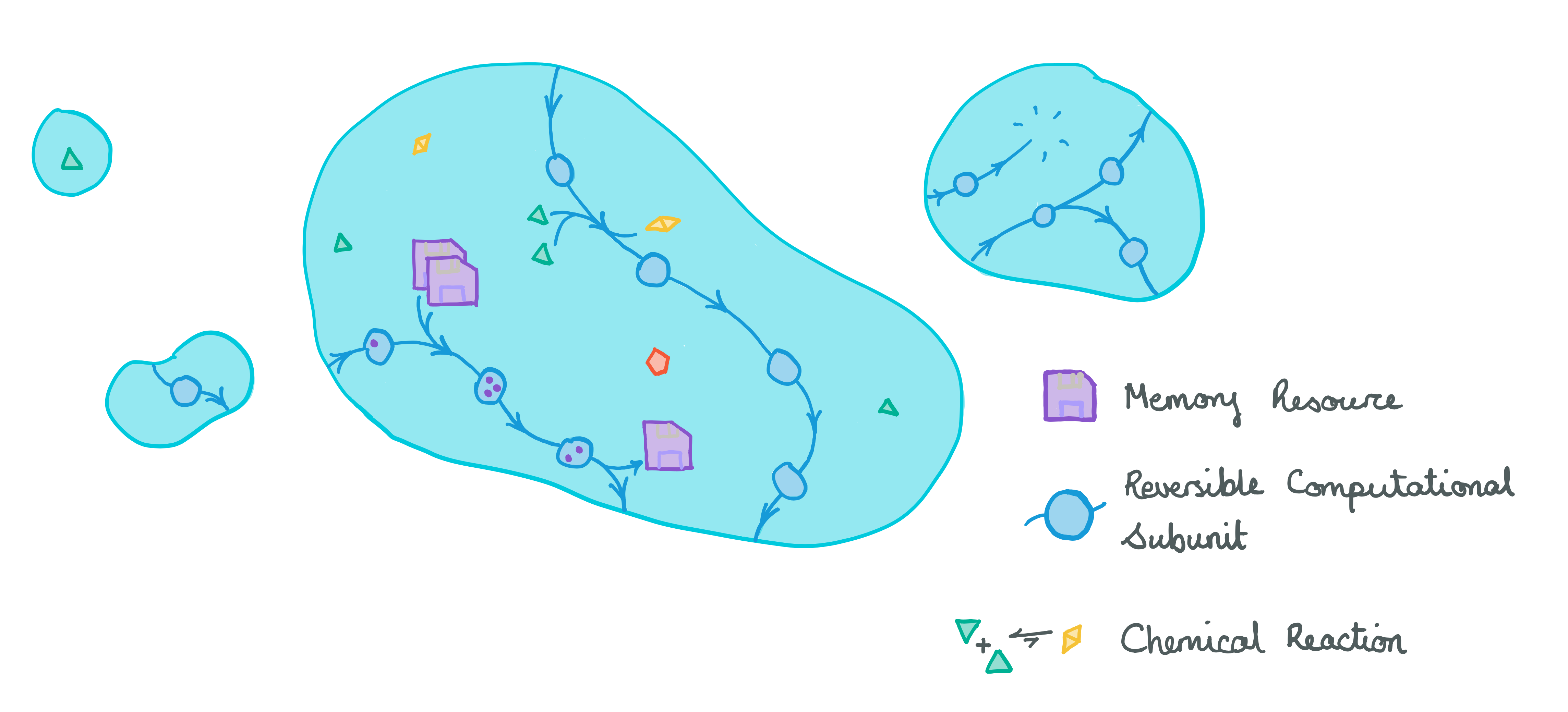}
    
  \caption{This paper concerns interactions between reversible Brownian computational particles and populations of other particles such as species involved in chemical reactions or shared memory resources. Populations of particles are subject to well known statistical mechanical laws, and these laws can cause problems when computational particles attempt to interact with these populations.}
  \label{fig:cover}
\end{figure}

\renewcommand{\abstractname}{Lay Summary}
\begin{abstract}
  In recent years, unconventional forms of computing ranging from molecular computers made out of DNA to quantum computers have started to be realised. Not only that, but they are becoming increasingly sophisticated and have a lot of potential to influence the future of computing. One interesting class of unconventional computers is that of reversible computers, which includes quantum computers. Reversible computing---wherein state transitions must be invertible, and therefore must conserve information---is largely neglected outside of quantum computing, but is a promising avenue for realising substantial gains in computational performance and energy efficiency.

  Reversible computing is of great interest because, unlike traditional irreversible computing which is subject to a lower bound on energy cost per operation~\cite{landauer-limit,szilard-engine}, there is no lower bound to the energy usage of a reversible computer. It might be imagined that one could go as far as to get \emph{processive}\footnote{That is, computation which proceeds forward at a finite, non-zero rate!}\ computation for free using reversibility, such as with the scheme of \textcite{fredkin-conlog}, but unfortunately mounting evidence~\cite{bennett-rev,frank-thesis} suggested otherwise. Indeed, in Part~I~\cite{earley-parsimony-i} of this series we proved that free processive computation is not possible, with instead the adiabatic regime being the best attainable. Still, for a given energy budget and spatial size these adiabatic reversible computers have asymptotically better performance scaling than irreversible computers subject to the same constraints.

  This analysis neglected, however, to consider interactions between subunits of these computers. In Part~II~\cite{earley-parsimony-ii} we analysed how computational subunits interact and communicate with one another. One may also wish to consider how computational subunits interact with other, non-computational subsystems, which in general will be governed by the laws of thermodynamics and statistical mechanics. When the free energy available to the computational subunits is limiting, this becomes a challenging endeavour as perturbations to these subsystems may have a corresponding entropy cost for which our available free energy is insufficient.

  Our Brownian computational subunits---themselves being small and particulate---will likely lack adequate internal memory for many programs of interest; as such, it can be expected that a realistic Brownian system will make heavy use of a shared pool of memory resources that can be dynamically distributed amongst the computational subunits according to their need. We therefore dedicate a section of this paper to analysing this particular case, and find that seemingly any possible scheme will not work unless additional free energy is supplied.

  The same conclusion applies to any chemical reaction we wish to drive away from equilibrium. In order to make these interactions viable, we propose an abstract chemical scheme to dynamically infer how much free energy is needed, and to supply it by accumulating just enough from the (weak) ambient free energy. The scheme takes any reaction, arbitrarily far from equilibrium, and sequesters some of the reactants and products such that the extant reactants and products are in equilibrium. The scheme rapidly adjusts to changes in the equilibrium position, and makes use of molecular computation for its inference algorithm. Unfortunately there is a cost in applying this scheme: the reaction is slowed by a factor inversely proportional to the `computational bias', $b\ll1$ (the net proportion of computational transitions that are successful). In fact, this overhead is comparable to that found for synchronisation interactions in Part~II.
\end{abstract}

\largertitlepage
\renewcommand{\abstractname}{Technical Abstract}
\begin{abstract}
  This paper concludes a three-Part series on the limits the laws of physics place on the sustained performance of reversible computers. Part~I concerned aggregate performance in terms of computational operations per unit time, but neglected to consider interactions among computational sub-units or between computational sub-units and shared resources such as memory or chemical species. Part~II extended the analysis to consider the former set of interactions. In this Part we extend the analysis to consider the latter set, with a particular focus on resource distribution in the first half. It is found that most schemes imaginable fail to function effectively in the limit of vanishing `computational bias' $b$, which measures the net fraction of transitions which are successful, and falls as the system grows in size. Driving thermodynamically unfavourable reactions, such as resource distribution, is a very general problem for such systems and can be solved by supplying a sufficient excess of free energy. We propose a scheme to dynamically supply enough free energy for a given reaction, automatically and rapidly adapting to changes in the disequilibrium state of said reaction---including the case when the favourable reaction direction switches. The overhead of this scheme is no worse than the overhead found in Part~II for communicating reversible computers under the same regime.
\end{abstract}

\clearpage

\section{Introduction}

This is the final Part of a three-Part series investigating the influence of the laws of physics on computation, particularly as they apply to large reversible Brownian computers (see \Cref{dfn:brownian}). It has long been known that reversible computation---computation which conserves information and whose every transition is invertible---allows for substantially better energy efficiency in principle. In particular, whilst irreversible computation is subject to a `Landauer limit' bounding the energy cost of each computational operation from below~\cite{landauer-limit}, reversible computation has no such lower bound. Not only can reversible computers achieve better energy efficiency, but their total computational rate can be shown to scale asymptotically faster with spatial size than irreversible computers occupying the same region (see \textcite{frank-thesis} for some early results). In fact, we proved~\cite{earley-parsimony-i} that the adiabatic regime of reversible computation is the best possible in any quantum or classical computer, in which the total computational rate (in terms of primitive computational operations per unit time) scales as $\sim\sqrt{AV}\sim R^{5/2}$ where $A$ is the convex bounding surface area of the computer, $V$ its volume, and $R$ its radius, compared to $\sim A$ for an irreversible computational system. Concretely, these are given by
\begin{align*}
  \frac4{\sqrt{3}}\pi R^{5/2}&\cdot\sqrt{\frac{\phi\lambda}{2k_BT}}
  &&\text{and}&
  4\pi R^2&\cdot\frac{\phi}{k_BT\Delta I},
\end{align*}
respectively for a spherical region, where $R$ is the radius of the sphere, $\phi$ is the power supply\footnote{The function of the power supply is to dissipate the entropy generated by the system, and thus the input power must have low entropic content in order to perform thermodynamic \emph{work}. Therefore, $\phi$ measures the rate of supply of \emph{free energy}, which is generally lower than the gross energy supply rate.} per unit surface area, $k_B$ is Boltzmann's constant, $T$ is the surface temperature, $\Delta I$ is the average amount of information erased per irreversible operation, and $\lambda$ is a rate per unit volume of gross reversible computation (whether forwards or backwards).

As discussed in Part II~\cite{earley-parsimony-ii}, the net computational rate is not the only performance metric of interest. Therefore, in Part II we evaluated computational performance of reversible Brownian computers (and argued for applicability of the results to reversible computers in general) whose individual computational components were permitted to interact with one another, i.e.\ to communicate and to perform concurrent/parallel computation. A key property of these interactions is that computational `particles' can be uniquely referenced, and therefore the traditional thermodynamic principles that apply to chemical systems were inapplicable. In this Part, we now consider the interface between traditional chemical species, of which there may be a variable number of identical units or particles, with the computational particles. For convenience, we shall refer to these as \emph{klona} and \emph{mona} respectively, following the formulation introduced in Part II. Recalling the definitions, we call particles which are uniquely distinguishable or addressable \emph{mona} (sg.\ \emph{monon}), after the Greek for unique, \emph{\gr μοναδικός}. Similarly, we call indistinct particles belonging to a species \emph{klona} (sg.\ \emph{klonon}). Klona, may be likened to traditional chemical species, such as glucose and ATP. Mona are less obvious; an example would be a \SI{150}{\milli\liter} solution containing a \SI{1}{\milli\molar} concentration of random nucleic acids of length 80 nucleotides\footnote{A nucleic acid, such as the DNA molecules that compromise our genome, is a polymer formed from an arbitrary sequence of four nucleotide monomers; random nucleic acids are easily obtained from DNA synthesis companies. As such, there are $4^{80}\approx\num{1.5e48}$ possible distinct 80-mers (nucleic acid of length 80). The birthday paradox says that we need $\num{1.4e24}$ of these random molecules before it is likely that there are duplicates present. Our \SI{150}{\milli\liter}, \SI{10}{\milli\molar} solution contains just shy of $\num{e20}$ such molecules, and it can be calculated that the probability that there are any duplicates in the solution is less than 1 in 350 million. Therefore, it is almost certain that every nucleic acid in said solution is a monon.}. We can therefore say that Part~II studies mona-mona interactions, and Part~III mona-klona interactions. We do not cover klona-klona interactions as these are well understood within the domains of chemistry and statistical physics.

Interactions between mona and klona are of great practical importance for Brownian computers; whilst mona by themselves may be capable of arbitrary computation, this computation only becomes useful when used to actuate some other system or when it otherwise produces some observable output. Examples may include the activation of a fluorophore upon reaching some logical condition, or directing the translation of a particular messenger RNA into a protein. Moreover, interaction with klona may be essential for computation: general computation requires access to an unbounded amount of memory, but by their very nature mona are limited in size and thus memory. Therefore it is important for mona to be able to recruit more memory as required, and release it for use by other mona when the need passes. Resources other than memory may also be considered, such as structural monomers for use in constructing and deconstructing molecular machines or computers.

\begin{dfn}[Brownian Computers]\label{dfn:brownian}
  A Brownian computer is a region of space (typically bounded or otherwise confined) containing particles. The definitions of particles and space are not important, and they may be complex macromolecules or simple elementary particles or even abstract entities in an abstract topological space. These particles are generally static in isolation, but when two or more particles collide they may alter their states during the interaction, in much the same way as a molecule of \ce{ATP} and a molecule of \ce{H2O} may collide, react, and beget the products \ce{ADP}, \ce{P_{i}} and \ce{H+} (particle number need not be conserved). These interactions must be \emph{reversible}, such that the products may collide and reform the reactants. To enable these interactions, particles must be endowed with velocities, and these velocities are assumed to follow a thermal distribution.

  A Brownian computer is characterised by being able to interpret the current state of the particles as a computational state: perhaps some particles are `computers' in their own right, and thus the evolution of their state over time is isomorphic to some model of computation, or perhaps the joint state of a number of particles must be considered. The former, wherein the computational particles are mona, is more powerful, whereas the latter approach, representing computational state with klona, typically results in the entire reaction volume being dedicated to a single program and thus limits the computational power. Multiple programs can be run if their components are orthogonal in the sense that they don't interact with components of a different program, but designing orthogonal klona is challenging and limited. An alternative mitigation approach is to use multiple bounded sub-volumes containing just the klona for that particular program; these sub-volumes may then be considered as abstract `particles' or mona, as can the combination of orthogonal sub-klona (forming a virtual sub-volume, if you will). Computational mona can be driven forward in computational phase space by a \emph{computational bias} (\Cref{dfn:bias}, below).

  An example would be that of molecular computers such as can be found in the field of DNA computing~\cite{rothemund-tm,winfree-tam,qian-seesaw,qian-sqrt,crn-rm,dsd,cardelli-dsd,qian-nn,winfree-dna-crn-univ,pen}. These consist of molecules of DNA dissolved in water, wherein the ability for nucleic acids to bind to other nucleic acids in a sequence-specific manner\footnote{Nucleic acids are polymers of arbitrary sequences of four monomers. In the case of DNA, these monomers are adenine, guanine, cytosine and thymine or \texttt{A}, \texttt{G}, \texttt{C} and \texttt{T}. The nucleotide monomers of nucleic acids are special in that they are subject to base-pairing rules: \texttt{A} preferentially binds to \texttt{T} and vice-versa, and \texttt{G} preferentially binds to \texttt{C} and vice-versa. The consequence is that a DNA sequence \texttt{5'-CTTGCATCGC-3'} will generally only bind stably to its (reverse) complementary sequence \texttt{5'-GCGATGCAAG-3'}.}\ is exploited in various different ways to program cascades of reactions encoding arbitrary computation. Those DNA computers using DNA-strand displacement as their mechanism have a klonal representation of computational state, whereas those using the Tile Assembly Model have a monal representation.
\end{dfn}

\begin{dfn}[Computational Bias]\label{dfn:bias}
  Recall from Part~I~\cite{earley-parsimony-i} that in a Brownian computer, transitions are mediated by systems of klona representing `positive' and `negative' bias. Physically, such klona act as a source of free energy and a real world example is given by $\ce{ATP}$ and $\ce{ADP + P_{\textrm i}}$ in biochemical systems. In the simplest case, there are $\oplus$ and $\ominus$ klona such that $\forall n.~\ce{$\ket{n}$ + $\oplus$ <=> $\ket{n+1}$ + $\ominus$}$ where $\ket{n}$ is the $n^{\text{th}}$ state of some monon, and this is positively biased when the concentration of $\oplus$ exceeds that of $\ominus$. We define the bias in this case to be $b=([\oplus]-[\ominus])/([\oplus]+[\ominus])$.
\end{dfn}

In this paper we shall be most interested in the case of resource distribution, and will consider various such schemes in \Cref{sec:resource-scheme}. One approach is a centralised store which mona can access, either by travelling there themselves or by sending a `courier' monon on their behalf to fetch and retrieve (or carry and return) a resource. This approach is undesirable. If the couriers float freely in solution, then reaching the central store and returning will both take a time $\sim\bigO{V}$; if on the other hand we introduce a lattice as in Part~II~\cite{earley-parsimony-ii}, then the average distance will be proportional to the radius $R$ and, given $b\sim R^{-1/2}$, the average time will be $\sim\bigO{b^{-3}}$. As the time for a single net computational step is $\sim\bigO{b^{-1}}$, both of these are untenable. Instead we shall seek a decentralised approach by making use of klona; intuitively it must become harder to access a resource as its scarcity increases, but we would like the timescale to do so to not be substantially more costly than $\sim (b\ce{[X]})^{-1}$ where $[\ce{X}]$ is the available concentration of the resource.

We will unfortunately find that the reactions involved in all possible schemes are unfavourable except in a very limited range of resource availability. In \Cref{sec:drive-unfav} we shall return to considering general mona-klona interactions, and demonstrate how mona can dynamically drive any unfavourable reaction with klona by inferring the number of bias tokens required to balance the entropic cost of the reaction. Alas, there will turn out to be an overhead in achieving this even in the case when the driven reaction is at equilibrium and hence cost-free.

\section{Resource Distribution Schemes}
\label{sec:resource-scheme}

{\def\figRRcap{An assortment of illustrative examples of some of the resource distribution schemes for a Brownian amorphous computer as considered in this paper.}
\begin{figure}[p]
  \centering
  \fbox{\begin{subfigure}{.72\linewidth}
    \centering
    \includegraphics[width=.88\linewidth]{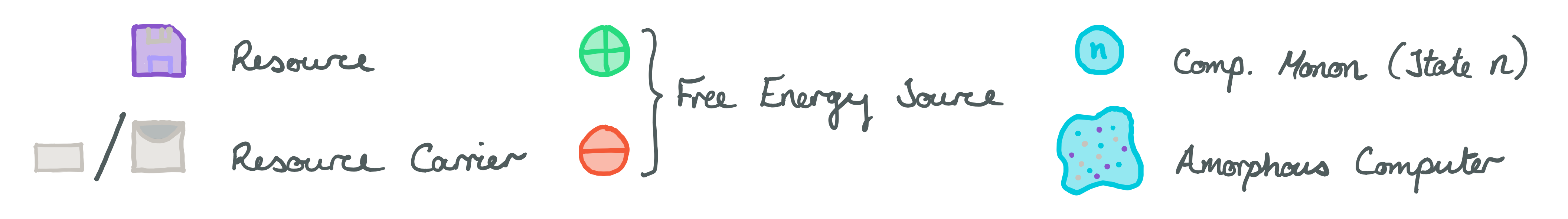}
  \end{subfigure}}\\[1em]
  \begin{subfigure}{.72\linewidth}
    \centering
    \includegraphics[width=.88\linewidth]{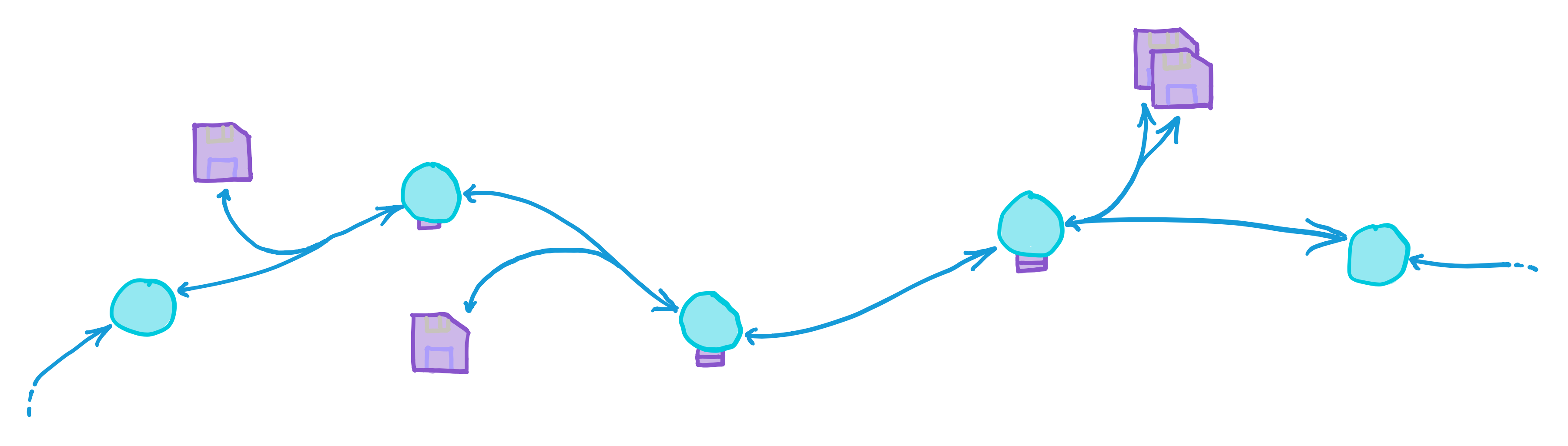}
    \caption{A possible state diagram for a monon interacting with a resource pool; the particular resource distribution scheme is left abstract.}\label{fig:rr-ex}
  \end{subfigure}\\[1em]
  \begin{subfigure}{.72\linewidth}
    \centering
    \includegraphics[width=\linewidth]{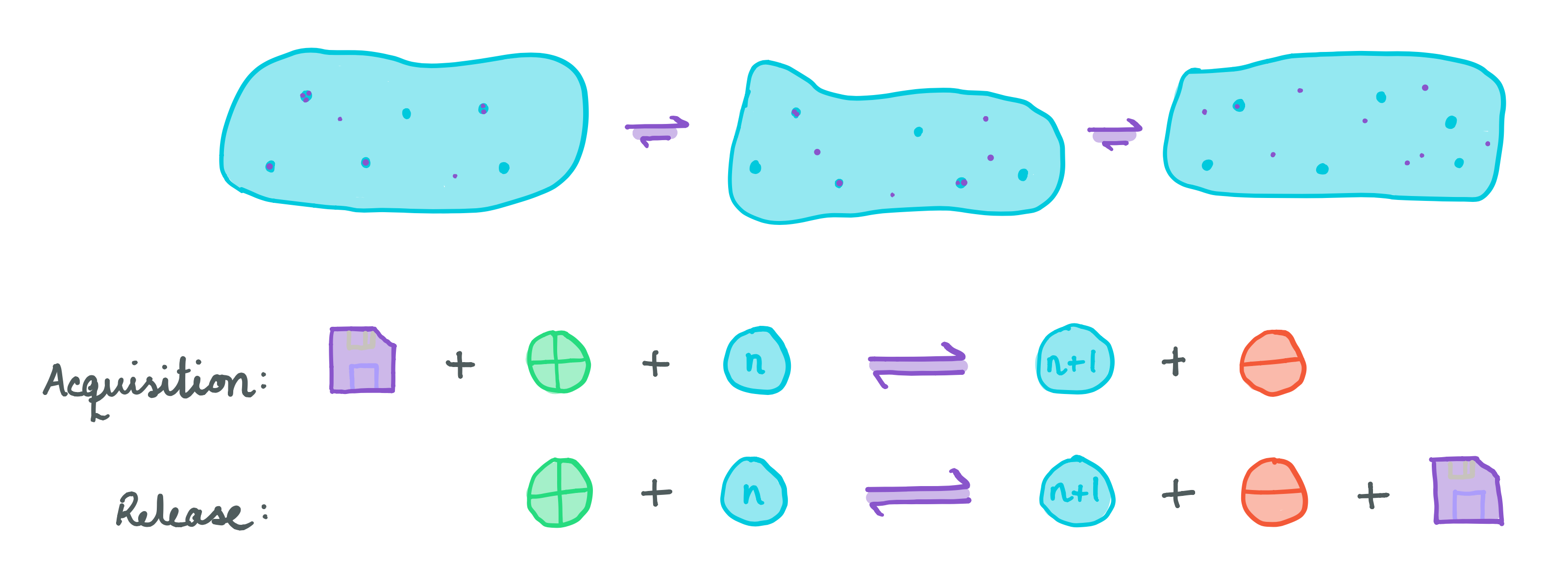}
    \caption{An illustration of the statistical states and reactions for the Free Klona resource distribution scheme.}\label{fig:rr-free}
  \end{subfigure}
  \caption{\figRRcap\ (continued)}\label{fig:rr}
\end{figure}
\begin{figure}[p]\ContinuedFloat
  \centering
  \fbox{\begin{subfigure}{.72\linewidth}
    \centering
    \includegraphics[width=.88\linewidth]{rr2-key}
  \end{subfigure}}\\[1em]
  \begin{subfigure}{.72\linewidth}
    \centering
    \includegraphics[width=\linewidth]{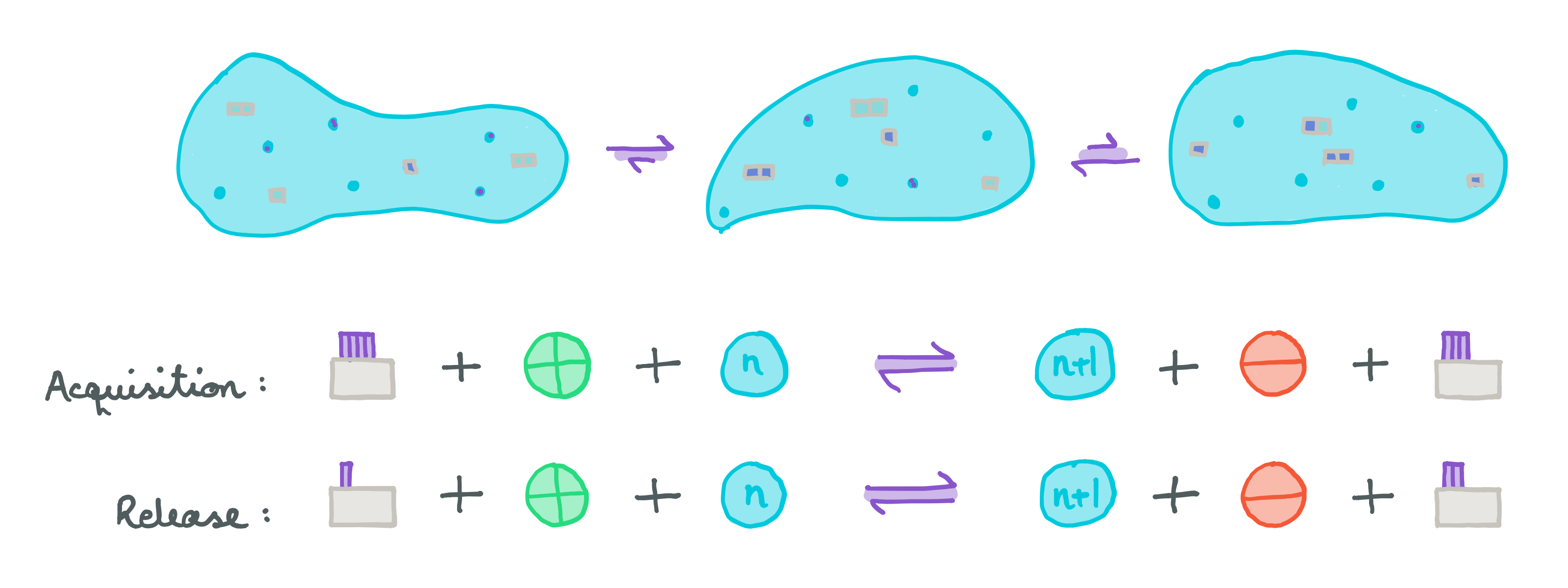}
    \caption{An illustration of the statistical states and reactions for the Bounded Carrier resource distribution scheme.}\label{fig:rr-bounded}
  \end{subfigure}\\[1em]
  \begin{subfigure}{.72\linewidth}
    \centering
    \includegraphics[width=\linewidth]{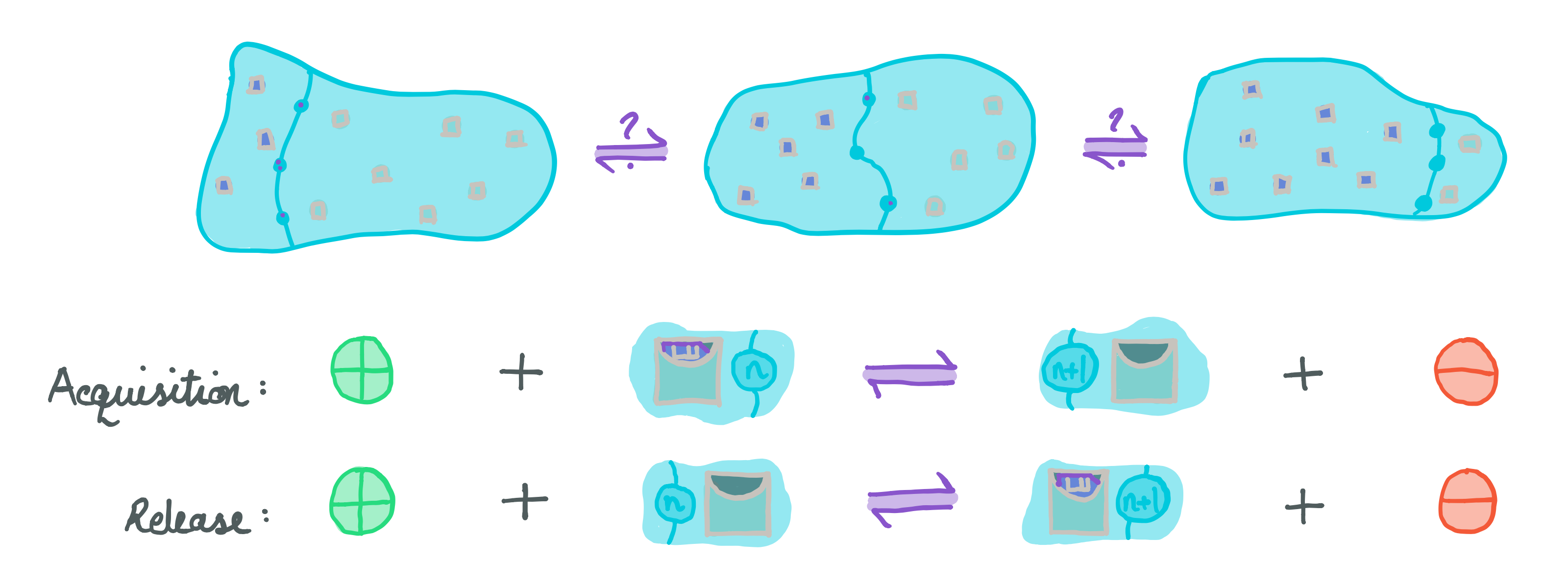}
    \caption{An illustration of the statistical states and reactions for the Isobaric Compartment resource distribution scheme. The question marks serve to indicate that the apparent isergonicity of the resource reactions is in question, and as we find, is unfortunately not the case.}\label{fig:rr-isobaric}
  \end{subfigure}
  \caption{(continued) \figRRcap}
\end{figure}}

A resource distribution scheme for a resource $\ce{X}$ is a set of klona from which $\ce{X}$ can be extracted and into which it can be deposited. Formally, resources are conserved discrete quantities analogous to charges in physics. Letting the set of resource species be indexed by $\mathcal R$ and the set of resource klona by $\Sigma$, then to each klonon $\sigma\in\Sigma$ is associated a resource charge $\vec q_\sigma\in\mathbb N^{\mathcal R}$: a vector over the field of naturals and indexed by the resource species. In any reaction, the net resource charge must be conserved; for example, for the reaction $\ce{ $\ket{n}$ + $\sum_{\sigma\in\Sigma}\nu_{\sigma}\sigma$ <=> $\ket{n+1}$ + $\sum_{\sigma\in\Sigma}\nu'_{\sigma}\sigma$ }$ involving the transition of a monon between states $\ket{n}$ and $\ket{n+1}$ and the interaction of said monon with resource klona according to the stoichiometries $\ce{ $\vec\nu$ <=> $\vec\nu'$ }$, this conservation law can be expressed as $\vec q_{\ket{n+1}} - \vec q_{\ket{n}} = \sum_{\sigma\in\Sigma} (\nu_\sigma-\nu_\sigma')\vec q_\sigma$.

\para{Free Klona}

The simplest possible scheme is that in which the resource is freely present in solution, as shown in \Cref{fig:rr-free}. That is, a system with $\mathcal R=\{\ce{X}\}$, $\Sigma=\{\ce{X}\}$ and $(\vec q_{\ce{X}})_{\ce{X}}=1$.
This system will be a viable resource pool so long as the free energy change involved in both extraction and deposition of $\ce{X}$ is less than the available computational free energy, $\log\frac pq=2\arctanh b$. Note that our notion of `free energy' uses a different convention than is normally taken, namely we are giving the change in information entropy of the universe $\Delta h$, whereas normally the free energy change is expressed as $\Delta G=-kT\Delta h$. The free energy change vanishes when the system is in equilibrium, which for $\ce{X}$ an ideal gas can be calculated (via the Sackur-Tetrode equation) to occur at a concentration
\begin{align*}
  [\ce{X}]_{(\text{eq.})} &= e^{5/2}\left(\frac{4\pi mu}{3h_P^2}\right)^{3/2},
\end{align*}
where $h_P$ is Planck's constant, $m$ is the mass of $\ce{X}$, and $u$ is its average internal energy. As we interact with the system, the concentration will be drawn away from this equilibrium position and as a result the absolute free energy difference will increase. For this scheme, the free energy difference corresponds to that due to adding or removing a particle of $\ce{X}$ and can therefore be identified with the \emph{Chemical Potential}, $\mu_{\ce{X}}$. In units of information, this can be expressed as $\mu_{\ce{X}}=\log\frac{[\ce{X}]\gamma_{\ce{X}}}{[\ce{X}]\gamma_{\ce{X}}|_{(\text{eq.})}}$ where $\gamma_{\ce{X}}$ is the activity coefficient of $\ce{X}$ that quantifies the deviation of the properties of $\ce{X}$ from an `ideal' substance. If $Q$ is the concentration of resource $\ce{X}$ (here trivially equal to $[\ce{X}]$) then we can now determine how far from equilibrium the resource system can be brought, and therefore how much of the resource $Q$ is accessible to the computational mona. For a small change $\delta Q$, we find
\begin{align*}
  \left|\frac{\delta Q}{Q}\right| < 2b\left|1+\pdv{\log\gamma_{\ce{X}}}{\log Q}\right|^{-1}  + \bigO{b^2}.
\end{align*}
We don't expect the activity coefficient to depend strongly on $Q$, and certainly not to the extent of $\pdv{\log\gamma}{\log Q}=-1$, which would imply a range of isentropic equilibria, and therefore we can only usefully access a proportion $\bigO{b}\ll1$ of the available resource $\ce{X}$ under this scheme.

\para{Bounded Carriers}

Perhaps the primary issue with free klona is that there is a change in particle number, implying a substantial entropy change when resources are traded with the pool. A possible improvement is suggested in ensuring total particle number remains constant by binding the resource to carrier klona, as shown in \Cref{fig:rr-bounded}. This may manifest, for example, as $\Sigma=\{\ce{X^0},{X^1}\}$ with $(\vec q_{\ce{X^0}})_{\ce{X}}=0$ and $(\vec q_{\ce{X^1}})_{\ce{X}}=1$. More generally, we have polymeric klona $\Sigma=\{\ce{X^k}:k\in[0,n]\cap\mathbb N\}$ with $(\vec q_{\ce{X^k}})_{\ce{X}}=k$ for some integer $n>1$. Once again, we shall consider a perturbation to the equilibrium distribution of the pool; it will be convenient to introduce a partition function $\mathcal Z(\beta)$ where the thermodynamic variable $\beta$ correlates with the amount of resource stored by the pool. The partition function will take the form $\mathcal Z=\sum_{k=0}^ne^{-\varepsilon_k-\beta k}$ where $\varepsilon_k$ contains all the thermodynamically pertinent information about $\ce{X^k}$ klona, such as its energy and internal degrees of freedom. In fact, we will take $\varepsilon_k=0$ as any $k$-dependence of the $\varepsilon_k$ would only serve to reduce the effective value of $n$ by narrowing the distribution\footnote{Assuming that the equilibrium distribution is unimodal, this corresponds to a reduction in variance. If the distribution is multimodal then instead there is a reduced `local' variance about each mode.}.

The free energy change for a reaction is given by the logarithm of the ratios of reaction rates (see, e.g., Section 4 of \textcite{earley-parsimony-i} for a derivation). Therefore in this case it is given by $\Delta h = \log\sum_{k=1}^n e^{-\beta k}-\log\sum_{k=0}^{n-1}e^{-\beta k}$,
and so at equilibrium (when $\Delta h=0$) we find $\beta=0$. For a perturbation from equilibrium, this reduces (exactly) to $\delta\Delta h = -\delta\beta$. Now, the average $\ce{X}$-charge of each klonon is given by $\bar k=-\partial_\beta\log\mathcal Z$, and the charge density by $Q=\sum_{k=0}^n k[\ce{X^k}]\equiv \bar k[\ce{X}]$. The proportional resource accessibility is then given by
\begin{align*}
  \left|\frac{\delta Q}{Q}\right| &= \left|\frac{-\delta\beta\var k}{\bar k}\right| < \tfrac13b(n+2)
\end{align*}
using $\bar k=\tfrac12n$ and $\var k=\tfrac1{12}n(n+2)$. This suggests taking $n\gtrsim b^{-1}$ to ensure full resource availability. Unfortunately, doing so would mean the total system charge scales as $V/b\sim V^{7/6}$ which clearly cannot be sustained in the limit of large systems. One potential resolution is to reduce the carrier concentration to $[\ce{X}]\sim b^{1/2}$ and then take $n\sim b^{-1/2}$ so that the total charge scales with $V$; whilst this would slow down resource reactions commensurately, mona could use this additional time to accumulate $\sim b^{-1/2}$ tokens in order to raise their free energy budget to $\Delta h\lesssim b^{1/2}$, thereby obtaining a resource accessibility of $|\delta Q/Q|\sim 1$ at a time penalty of $\sim b^{-3/2}$.

\para{Unbounded Carriers}

It is also possible to consider polymeric klona of unbounded size, i.e.\ the limit $n\to\infty$, in which case the partition function is given by $\mathcal Z=(1-e^{-\beta})^{-1}$. It should be noted that in this limit there is no equilibrium distribution of the resources, as $\mathcal Z$ diverges as $\beta\to0$. In fact, distributions only exist for $\beta>0$ whilst the free energy constraint bounds it from above and hence $0<\beta<2b$. This range corresponds to $\bar k>(2b)^{-2}-\bigO{1}$ and therefore a total enclosed charge $\gtrsim V^{4/3}$; for lower densities, the only favourable reaction is resource deposition. Therefore, whilst conceptually simpler, unbounded carriers are to be avoided.

\para{Isobaric Compartments}

The problem inherent to seemingly all schemes is that a resource reaction must necessarily reduce the quantity of reactants and increase the quantity of products, and this will in turn bias the resource reactions in the opposite direction per \emph{Le Chatelier}'s principle. As our last attempt to circumvent these issues, we introduce a scheme exploiting varying partial volumes in order to maintain equal concentrations of reactants and products. This `Isobaric Compartment' scheme is illustrated in \Cref{fig:rr-isobaric} and consists of dividing the system into two compartments by means of a massless and infinitely flexible membrane, impermeable to the carrier species $\ce{X^0}$ and $\ce{X^1}$. Referring to the two compartments by $\Gamma_\oplus$ and $\Gamma_\ominus$, we then establish the invariants that $\Gamma_\oplus$ only contains $\ce{X^1}$ and $\Gamma_\ominus$ only contains $\ce{X^0}$. A monon wishing to exchange resources with the pool must first bind to the membrane, and then to maintain the invariant it will translocate the carrier species between the two compartments. At equilibrium, the concentrations of carrier species in each compartment must be equal and thus it would appear that we have been successful in eliminating the free energy cost for this mona-klona interaction.

\newcommand{\eff}{{\text{eff.}}}
\newcommand{\eqm}{{\text{eq.}}}
Unfortunately, whilst the carrier species concentrations are equal, concentration is not the correct quantity for determining the reaction rates and thence the free energy difference. From a collision theory perspective, the reaction rates are given by the frequency with which mona and klona collide (in the correct orientation and conformation). Consider the case where all but one carrier reside within $\Gamma_\oplus$: if the total number of carriers is $N$ then the mean volume of $\Gamma_\ominus$ will be $V/N$ and the mean concentration $1/(V/N)=N/V$. This would suggest that a monon bound to the interfacial membrane will encounter the $\ce{X^0}$ klonon as often as it does an $\ce{X^1}$ klonon, but this clearly cannot be the case. If we imagine the $\ce{X^0}$ klonon as residing within a small bubble of volume $V/N$, then whilst its local concentration is indeed the same as for the $\ce{X^1}$ klona, the bubble itself is free to migrate anywhere along the membrane. Consequently, in a system in which particles can bind to the boundary of the system in addition to exploring their volume, the effective concentration becomes $N/(\alpha + \gamma)$ where $\alpha$ is the effective `volume' of the boundary (proportional to its area) and $\gamma$ is the enclosed conventional volume. For our isobaric system, $\gamma_\oplus=N_\oplus\gamma_0$ (and similarly for $\gamma_\ominus$) where the constant $\gamma_0=V/(N_\oplus+N_\ominus)$ is the volume available to each particle. We can therefore write $[\ce{X^0}]_\eff=[\ce{X^0}]/(\alpha/V + [\ce{X^0}]\gamma_0)$ where $[\ce{X^0}]=N(\ce{X^0})/V$ is the `true' concentration; redefining terms, we let $[\ce{X^0}]_\eff=[\ce{X^0}]/(\alpha + [\ce{X^0}]\gamma)$.

The system can be generalised to carriers for $n>1$ by letting the membrane be semi-permeable to $\{\ce{X^k}:0<k<n\}$, as these klona can participate as both reactants and products in either reaction. We can then write $\mathcal Z_\ominus=\sum_{k=0}^{n-1}e^{-\beta_\ominus k}$ for the partition function of $\Gamma_\ominus$ and similarly for $\Gamma_\oplus$. The effective concentrations of $\ce{X^k}$ in each compartment must be equal for $0<k<n$, i.e.\ $(e^{-\beta_\ominus k}/\mathcal Z_\ominus)[\ce{X}]_{\ominus\eff}=(e^{-\beta_\oplus k}/\mathcal Z_\oplus)[\ce{X}]_{\oplus\eff}$. As this must apply for each $0<k<n$, $(\beta_\oplus-\beta_\ominus)k$ must be constant with respect to $k$ and therefore $\beta_\oplus=\beta_\ominus=\beta$. Moreover, $[\ce{X}]_{\oplus\eff}/[\ce{X}]_{\ominus\eff}=\mathcal Z_\oplus/\mathcal Z_\ominus=e^{-\beta}$. Equilibrium occurs once again for $\beta=0$ with corresponding `true' concentrations $[\ce{X}]_\oplus=[\ce{X}]_\ominus=\tfrac12[\ce{X}]_{\text{total}}$. The free energy difference is then given by
\[\delta\Delta h=-\delta\beta=\frac{\delta[\ce{X}]_\oplus}{\tfrac12[\ce{X}]}\frac{2\alpha}{\alpha+\tfrac12\gamma[\ce{X}]}.\]
Finally, the resource charge is $Q=\bar k_\oplus[\ce{X}]_\oplus+\bar k_\ominus[\ce{X}]_\ominus$ and hence the proportional availability is given by
\begin{align*}
  \left|\frac{\delta Q}{Q}\right| &= \left|\frac{-\delta\beta\tfrac12[\ce{X}](\var_\oplus k+\var_\ominus k) + \delta[\ce{X}]_\oplus}{\tfrac12n[\ce{X}]}\right| \\
  &< 2b \left(\frac{\tfrac16(n^2-1)}{n} + \frac{\alpha+\tfrac12\gamma[\ce{X}]}{2\alpha n}\right) \\
  &< b ( \tfrac13n+\tfrac1{6n}  + \tfrac{1}{4\alpha n} ) ,
\end{align*}
recalling that $\gamma\equiv1/[\ce{X}]$.

As with the Bounded Carrier scheme we can let $n\sim1/b$ to ensure full resource availability, but we can also choose to let $\alpha\sim b$ with $n=1$. The constant $\alpha$ is proportional to the ratio of the membrane area to the system volume. A naive implementation would of course have $\alpha\sim b^2$ which, whilst ensuring full resource availability, would limit the number of mona able to interact with the resource scheme proportional to $A$. Instead, consider replacing the system's volume with a space-filling tubule (or perhaps a tubule-lattice) and partitioning the tubule along its length; this system is still an instance of the Isobaric Compartment scheme, but allows an $\alpha$ as large as $\sim1$ (if we wanted to allow all mona to interact with the resource system simultaneously). By making the tubule diameter $\sim b^{-1}$, a value of $\alpha\sim b$ can be achieved in which case the proportion of mona able to simultaneously interact with the resource system will be $\sim V^{5/6}$. This is, however, a time penalty of $\sim b^{-2}$ and hence is strictly worse than the Bounded Carrier scheme.

\section{Driving Unfavourable Reactions}
\label{sec:drive-unfav}

We conjecture that the results in the previous section are general; that is, no resource scheme can achieve better resource availability under the same constraints. In order to make use of a greater proportion of the resource pool, more free energy must be supplied and, ideally, the amount of free energy supplied should be the minimum necessary to avoid wasting the supply of negentropy. As alluded to in the introduction, we shall now consider general mona-klona interactions of arbitrary free energy cost.

\para{Reactions of Known or Bounded Cost}

Consider an arbitrary reaction $\ce{ $\sum_{i}\nu_{i}$X_i <=> $\sum_{i}\nu'_{i}$X_i }$ with forward reaction rate $\alpha$ and backward rate $\beta$. Recalling that the free energy change for the forward reaction is given by $\Delta h=\log\frac\alpha\beta$, it can be seen that $n>-\Delta h/2\arctanh b$ computational bias tokens are needed to drive the reaction forward in the case $\Delta h<0$; similarly, if $\Delta h>0$ then $n>\Delta h/2\arctanh b$ tokens are needed to drive the reaction backward. In addition, if we wish to couple such an unfavourable reaction to our computational mona then we should ensure by some means that the reaction does not occur in uncoupled isolation, such as raising the activation energy of the reaction or by employing a reaction mechanism that requires interaction with the mona. This kind of coupling abounds in nature, wherein a variety of bias systems are exploited. In fact, the $\ce{ATP}:\ce{ADP + P_i}$ bias system is itself generated from a number of other bias sources. One particular example involves the conversion of the free energy of a transmembrane proton gradient for which $n\gtrsim 4$; the molecular machine responsible for this coupling, ATP synthase, is a molecular motor which translocates 11 protons\footnote{The exact number varies between species, and even between different organelles of the same species, but is typically between 10 and 14.} across the membrane per revolution, in the process converting 3 molecules of $\ce{ADP}$ to $\ce{ATP}$, an entropically unfavourable process.

When $n$ is known and fixed (or, at least, bounded from above) an optimal approach is provided by the simple scheme that performs $n$ transitions between each unfavourable reaction. Indexing the mona states, $\ce{M_k}$, modulo $n+1$, the scheme can be written thus
\begin{align*}\begin{gathered}
  \ce{M_0 <=>>[$p\lambda$][$q\lambda$] $\cdots$ <=>>[$p\lambda$][$q\lambda$] M_k <=>>[$p\lambda$][$q\lambda$] $\cdots$ <=>>[$p\lambda$][$q\lambda$] M_n},\\
  \ce{M_n + $\sum_{i}\nu_{i}$X_i <=>>[$\alpha'$][$\beta'$] M_0 + $\sum_{i}\nu'_{i}$X_i}.
\end{gathered}\end{align*}
The reactions between $\ce{M_k}$ and $\ce{M_{k+1}}$ serve to increase the ratio $[\ce{M_n}]/[\ce{M_0}]$ such that it counteracts the ratio $\alpha'/\beta'=\alpha/\beta<1$. To solve for the steady state dynamics of this system, we write the currents
\begin{align*}
  S(\ce{M_k}\mapsto\ce{M_{k+1}}) &= p\lambda[\ce{M_k}] - q\lambda[\ce{M_{k+1}}] \\
  S(\ce{M_n}\mapsto\ce{M_0}) &= \alpha'[\ce{M_n}] - \beta'[\ce{M_0}]
\end{align*}
and use the fact that, at steady state, the currents must be equal. Moreover, the current will correspond to the net rate at which the target reaction is driven, and therefore we can find the optimal choice of $n>|\Delta h/2b|$ to maximise this rate (as for excessively large $n$, the chain of intermediate monon states will dilute the proportion of transitions performing the desired reaction).

We first show that the system makes optimal use of the consumed bias tokens by calculating the value of $n$ for which the current vanishes, and hence for which the reaction is exactly balanced. For vanishing current, the principle of detailed balance applies and the steady state distribution is obtained simply by \smash{$[\ce{M_k}]/[\ce{M_0}]=(\frac pq)^k$} and \smash{$[\ce{M_0}]/[\ce{M_n}]=\alpha/\beta$}. This yields \smash{$n=\log\frac\beta\alpha/\log\frac pq=-\Delta h/2\arctanh b$}, precisely the threshold value of $n$. The attentive reader may point out that this value of $n$ is not necessarily integral, and therefore a cyclic chain of length $n+1$ may not exist; this is true, but we shall be seeking larger values of $n$ to increase the rate of the reaction and for which picking the nearest integral value shall be acceptable. Alternatively, one could consider the superposition of two such chains for $\lfloor n\rfloor$ and $\lceil n\rceil$ in the appropriate proportions.

To solve for non-vanishing current, we perform two telescopic sums
\begin{align*}
  nS &= \sum_{k=0}^{n-1} S(\ce{M_k}\mapsto\ce{M_{k+1}})  &
  \frac1pS \sum_{k=0}^{n-1} t^{-k} &= \sum_{k=0}^{n-1}( t^{-k} \lambda[\ce{M_k}] - t^{-k-1} \lambda[\ce{M_{k+1}}] )  \\
  &= b\lambda[\ce{M}] + q\lambda[\ce{M_0}] - p\lambda[\ce{M_n}], &
  S \frac{p^n-q^n}{p-q} &= p^n\lambda[\ce{M_0}] - q^n\lambda[\ce{M_n}],
\end{align*}
where $t\equiv\frac pq$, and then substitute $S(\ce{M_n}\mapsto\ce{M_0})$ for the current:
\begin{align*}\begin{gathered}
  b\lambda[\ce{M}] = (p\lambda+n\alpha')[\ce{M_n}] - (q\lambda+n\beta')[\ce{M_0}], \\
  \left(q^n\lambda + \frac{p^n-q^n}{p-q}\alpha'\right)[\ce{M_n}] = \left(p^n\lambda + \frac{p^n-q^n}{p-q}\beta'\right)[\ce{M_0}].
\end{gathered}\end{align*}
Hence, solving for $[\ce{M_0}]$ and $[\ce{M_n}]$, we can obtain the steady state current thus:
\begin{align*}
  S &= b\lambda[\ce{M}] \frac{q^n\beta' - p^n\alpha'}{(q^{n+1}-p^{n+1})\lambda + n(q^n\beta'-p^n\alpha')+(q\alpha'-p\beta')\frac1b(p^n-q^n)}.
\end{align*}
After some rearrangement, we find that the current is maximised when
\begin{align*}
  (t^{n/2}-t^{-n/2}\gamma)^2 &= \frac{q\log t}{b}(\gamma-1)(t\gamma-1)+\frac{q\lambda\log t}{\alpha}(t\gamma+1)
\end{align*}
where $\gamma\equiv\beta/\alpha$. For small bias, this reduces to $n=2n_0 + \bigO{b}$ where $n_0=|\Delta h|/2\arctanh b$ is the threshold value. When $n$ is large, this gives a current $S \to [\ce{M}]\min(2b^2\lambda\tfrac\alpha\beta, 2b\alpha)$. This assumes that the intermediate monon transitions are not useful; if the intermediate transitions in fact perform useful work then this current can be multiplied by $n$ and, in the limit of large $n$, the current tends to $b\lambda$: the net transition rate for uncoupled mona.

\para{Reactions of Unknown or Unbounded Cost}

The preceding scheme suffers from a number of shortcomings due to the fact that $n$ is fixed; when $n_0$ is lower than that designed for, the system expends more bias tokens than necessary, whereas when it is greater the system will stall. Moreover the system is asymmetric with respect to direction of the driven reaction: if $\beta>\alpha$ then the forward reaction should be driven with $n$ tokens whilst the reverse reaction needs no additional bias, but this complicates the act of running a subroutine in reverse as each bias chain for the forward reaction must be removed, and a bias chain must be introduced for each instance of the reverse reaction. This is further complicated if the intermediate transitions along the chain perform useful computation. Whilst these problems are not insurmountable, it is possible to solve them all with a revised scheme in which $n$ is permitted to vary.

A scheme with variable $n$ is incompatible with the specific implementation of that for fixed $n$, and so we first migrate the mona chains into a distinct system of klona consisting of the species $\{\ce{X}^{(n)}_{k}:n,k\in\mathbb N\land 0\le k\le n\} \cup \{\ce{X}^{(-n)}_{-k}:n,k\in\mathbb N\land 0\le k\le n\}$ and subject to reactions \smash{$\ce{ X$^{(n)}_{k}$ <=>>[$p$][$q$] X$^{(n)}_{k+1}$ }$}, with steady state $[\ce{X}^{(n)}_{k}]/[\ce{X}^{(n)}_{0}]=(\frac pq)^k$. To demonstrate practicability, an abstract molecular realisation is provided in \Cref{app:seq-klona}. The idea is to sequester some of the reactants and products in an inactive state, with those remaining in an active state being at equilibrium (for an isenthalpic reaction, this manifests as equal concentrations). Representing the reactants by $\ce{A}$ and the products by $\ce{B}$, this effective sequestration behaviour is achieved by the reaction $\ce{ M$_i$ + A{:}X$^{(n_0)}_{n_0}$ <=>>[$p$][$q$] M$_{i+1}$ + B{:}X$^{(n_0)}_0$ }$ with $n_0=-\Delta h/2\arctanh b$. For fixed $n_0$, this scheme already solves all the aforementioned problems as the complexed system \smash{$\ce{A{:}X$^{(n_0)}_{n_0}$ <=> B{:}X$^{(n_0)}_0$}$} is an equilibrated abstraction over the underlying $\ce{A <=> B}$ reaction, and hence can be coupled directly to the computational mona system.

In the course of system operation, the value of $n_0$ will almost certainly vary. Introducing the sequestration klona for all values of $n\in\mathbb Z$, and letting $\alpha' \equiv \alpha\sum[\ce{ X$^{(n)}_n$ }]$ and $\beta' \equiv \beta\sum[\ce{ X$^{(n)}_0$ }]$, we see that the desired equilibrium condition is given by $\alpha'=\beta'$. Therefore, if the values of $\alpha'$ and $\beta'$ deviate from equilibrium then we should counteract this deviation by perturbing the $n$-distribution of the sequestration klona. Ideal kinetics are attained if this $n$-distribution converges to $\alpha'=\beta'$ exponentially fast, i.e.\ $\dot n\propto \beta'-\alpha'$. This target kinetics arises because $[\ce{ X$^{(n)}_n$ }]=(\frac pq)^n[\ce{ X$^{(n)}_0$ }]$ and $\frac pq>1$, hence increasing $n$ increases $\alpha'/\beta'$. A first attempt at implementation might be given by $\forall n\in\mathbb Z. \ce{ M$_i$ + A{:}X$^{(n)}_{n}$ <=>>[$p$][$q$] M$_{i+1}$ + B{:}X$^{(n-1)}_0$ }$, but this doesn't quite replicate the desired kinetics. Instead we separate the concerns of the computational bias and the reaction coupling thus:
  \begin{equation}\begin{aligned}
    {\longce\begin{rcases}
      \ce{ M_i <=>[$p$][$p$] M$_{i+\frac13}$ } \\
      \ce{ M_i <=>>[$p$][$q$] M$_{i+\frac13}$ }
    \end{rcases}} &&
    \ce{ M$_{i+\frac13}$ + A{:}X^{(n)}_{n} <=> M$_{i+\frac23}$ + B{:}X^{(n-1)}_0 } && 
    {\longce\begin{cases}
      \ce{ M$_{i+\frac23}$ <=>[$q$][$q$] M_{i+1} } \\
      \ce{ M$_{i+\frac23}$ <=>>[$p$][$q$] M_{i+1} }
    \end{cases}}
  \end{aligned}\label{eqn:dyn-coupling-scheme}\end{equation}
Assuming that at steady state $[\ce{M$_{i+\frac13}$}]=[\ce{M$_{i+\frac23}$}]\equiv\mu$, the coupled reaction then replicates the kinetics $\dot n=k\mu(\beta' - \alpha')$. For symmetry, we have introduced two intermediate mona with their inter-monon transitions used for coupling to the computational bias. In order to ensure that only a single token's worth of bias is consumed in the reaction, we also introduce two leak reactions such that the total free energy change is precisely that for a single token: $\Delta h=\log\frac{p+p}{q+p} + \log\frac{p+q}{q+q}=2\arctanh b$.

To verify that this system has the desired properties, we now solve for the steady state distribution. Employing detailed balance, we find for each $n$ that $[\ce{X$^{(n)}_{n}$}]/[\ce{X$^{(n)}_{n}$}]$
\begin{align*}
  \frac{[\ce{X$^{(n)}_{n}$}]}{[\ce{X$^{(n-1)}_{0}$}]} &= \frac\beta\alpha \equiv \left( \frac pq \right)^{n_0}  & &\implies &
  \frac{[\ce{X$^{(n)}_{0}$}]}{[\ce{X$^{(0)}_{0}$}]} &= \left( \frac pq \right)^{-\frac12(n+\frac12-n_0)^2+\frac12(\frac12-n_0)^2},
\end{align*}
where we have used the intra-$\ce{X$^{(n)}_{k}$}$ steady state distribution, $[\ce{X}^{(n)}_{k}]/[\ce{X}^{(n)}_{0}]=(\frac pq)^k$. This steady state distribution resembles a discrete Gaussian in $n$, and hence is relatively tightly centered about $n_0$; this is important to ensure good control over the distribution and that perturbing its mean value is practicable. As a sanity check, we can show that this distribution indeed gives $\beta'=\alpha'$:
\begin{align*}
  \frac{\beta'}{\alpha'} &= \left( \frac pq \right)^{n_0} \frac{\sum_{n\in\mathbb Z}( \frac pq )^{-\frac12(n+\frac12-n_0)^2+\frac12(\frac12-n_0)^2}}{\sum_{n\in\mathbb Z}( \frac pq )^{n}( \frac pq )^{-\frac12(n+\frac12-n_0)^2+\frac12(\frac12-n_0)^2}} \\
  &= \frac{\sum_{n\in\mathbb Z}( \frac pq )^{-\frac12(n+\frac12-n_0)^2+\frac12(\frac12+n_0)^2}}{\sum_{n\in\mathbb Z}( \frac pq )^{-\frac12(n-\frac12-n_0)^2+\frac12(\frac12+n_0)^2}} \\
  &= 1.
\end{align*}
Finally we seek to characterise the rates for the coupled reaction; first, we shall require the weight for each of the $\{\ce{X$^{(n)}_{k}$}:k\}$ subsystems:
\begin{align*}
  \sum_{k=0}^n[\ce{X$^{(n)}_{k}$}] &= \frac{[\ce{X$^{(n)}_{0}$}]}{t-1}\begin{cases}
    t^{n+1}-1 & n \ge 0 \\
    t-t^{n} & n \le 0
  \end{cases} 
\end{align*}
where $t=\frac pq$. Next, we use $t^nt^{-\frac12(n+\frac12-n_0)^2+\frac12(\frac12-n_0)^2}=t^{-\frac12(n-\frac12-n_0)^2+\frac12(\frac12+n_0)^2}$ to find the normalisation,
\begin{align*}
  [\ce{X$^{(0)}_{0}$}]^{-1} = 1 &+ \textstyle\frac1{t-1}[t\sum_{n<0}-\sum_{n>0}\,]t^{-\frac12(n+\frac12-n_0)^2+\frac12(\frac12-n_0)^2} \\
  &+ \textstyle\frac1{t-1}[t\sum_{n>0}-\sum_{n<0}\,]t^{-\frac12(n-\frac12-n_0)^2+\frac12(\frac12+n_0)^2}.
\end{align*}
This is non-trivial for general $n_0$, so we restrict to the two limiting cases of $n_0=0$ and $|n_0|\gg1$. In the former case, the normalisation simplifies to $[\ce{X$^{(0)}_{0}$}]=\frac{b}{1+b}$ and so we have $\beta'=\alpha'=\frac{\alpha b}{1+b}\sum_{n\in\mathbb Z}t^{\frac18-\frac12(n-\frac12)^2}$. The sum is bounded from below by the integral $2\int_0^\infty t^{-\frac12(s+\frac12)^2}\dd{s}$, and from above by $2\int_0^\infty t^{-\frac12(s-\frac12)^2}\dd{s}$, from which we obtain $\alpha'=\alpha\sqrt{\pi b}+\bigO{b}$. In the latter case, suppose $n_0\gg1$ is positive (i.e.\ $\beta\gg\alpha$). The normalisation will then be dominated by the sums over $n>0$ and so has asymptotically exact approximation
\begin{align*}
  [\ce{X$^{(0)}_{0}$}]^{-1} &=\textstyle \frac{t}{t-1}\sum_{n\in\mathbb Z}t^{-\frac12(n-\frac12-n_0)^2+\frac12(\frac12+n_0)^2}-\frac1{t-1}\sum_{n\in\mathbb Z}t^{-\frac12(n+\frac12-n_0)^2+\frac12(\frac12-n_0)^2} \\
  &=\textstyle \frac 1b[pt^{\frac12(\frac12+n_0)^2}-qt^{\frac12(\frac12-n_0)^2}]\sum_{n\in\mathbb Z}t^{-\frac12(n+\frac12-n_0)^2},
\end{align*}
hence $\alpha'$ reduces to $2b\alpha + \bigO{b^2}$ in the case of $\beta\gg\alpha$, or $2b\beta + \bigO{b^2}$ in the case of $\alpha\gg\beta$. In the ideal case, in which only the excess $\ce{B}$ (when $\beta>\alpha$) is sequestered, we should have $\alpha'=\beta'=\alpha$. The consequence of the factor $(\sqrt{\pi b},2b)$ is to further suppress the current through the coupled transition. As the uncoupled monon transitions have current $\propto b$, this results in an overhead of $\sim b^{-3/2}$ in the best case and $\sim b^{-2}$ in the worst. This overhead is comparable to the time penalties we found for communication and synchronisation interactions in Part~II~\cite{earley-parsimony-ii}, as for the coupled reaction to proceed both the monon and the complexed klonon must be in receptive states. Drawing further from the analogy, one may wonder whether reactions involving multiple klona are subject to even greater overheads; fortunately, this is not the case: provided only one sequestration klonon participates, by choosing a single representative from each of the reactant klona and product klona to engage in complexation with the sequestration klona, then the results stand.

\section{Conclusion}

In this paper we studied the interaction of mona, such as molecular computers, with klona, shared resources, in a system with a limiting supply of free energy. As in Part~II~\cite{earley-parsimony-ii}, we found this to be very expensive in comparison to independent computation. In general the cost, in terms of transition time, to couple mona to an arbitrary system of klona is of order $\bigO{b^{-2}}$ (an overhead of $\bigO{b^{-1}}$) but, in contrast to the mona-mona interactions of Part~II, lighter overheads of order $\bigO{b^{-1/2}}$ or better can be achieved under certain conditions. For example, whilst we found that all resource distribution schemes we could devise (some particular examples of which were illustrated in \Cref{sec:resource-scheme}) did not admit a fractional resource availability above $\bigO{b}$ if time overheads were to be avoided, by allowing a time overhead of $\bigO{b^{-1/2}}$ (thence a time penalty of $\sim\bigO{b^{-3/2}}$) the full resource pool was rendered accessible. We conjectured that these results are in fact general bounds on the capabilities of any such system.

To drive a general out-of-equilibrium system of klona, one needs to supply sufficient free energy to overcome the entropy cost of the reaction. When this cost is known or bounded, such as in biological systems, the definition of `sufficient' can be precomputed and built in to the specifications of the system. In general, however, the cost may be unknown or unbounded, or the favourable direction of the reaction may switch over time. We proposed a scheme in \Cref{sec:drive-unfav} to dynamically infer this cost, as well as automatically applying the correct amount of free energy tokens to pay for it, thereby exposing an equilibrated interface to the disequilibrium reaction. A more detailed specification of the (abstract) molecular realisation of such a scheme is presented in \Cref{app:seq-klona}. This equilibrated interface can then be readily coupled to our monon transitions, however the cost must still be paid in terms of transition time overhead and indeed it is, as the concentration of the interface is commensurately reduced. Unfortunately our proposed scheme has a minimum overhead of $\bigO{b^{-1/2}}$ even in the case where the underlying reaction is at or near equilibrium (where it is theoretically possible to achieve zero overhead). It is unknown whether a better scheme exists, whilst still retaining the ability to drive reactions that are arbitrarily far from equilibrium.

As any non-zero overhead to a class of interactions will lead to this class `freezing out' in very large reversible computers, their frequency must be minimised---the proposed scheme notwithstanding. Once again referencing the conclusions of Part~II, these interactions thus minimised can make use of a subpopulation bias klona of greater free energy to drive these unfavourable interactions more effectively.

\appendix

\section{Acknowledgements}

The author would like to acknowledge the invaluable help and support of his supervisor, Gos Micklem. This work was supported by the Engineering and Physical Sciences Research Council, project reference 1781682.

\section{Molecular Counter Implementation}
\label{app:seq-klona}

Here we present example implementations of the counter klona $\ce{X^{(n)}_k}$ used in \Cref{sec:drive-unfav}. These implementations are abstract, but point towards a viable molecular realisation.

\subsection{Unary Representation}

\begin{listing}
  \centering
  {\longce\fbox{\begin{minipage}{.85\textwidth}\begin{align*}
    \ce{ $\llb\ctrTok\sigma\mkrAnyR\tokSucc\ctrTok\sigma'\rrb$ &<=>>[$p$][$q$] $\llb\ctrTok\sigma\tokSucc\mkrAnyR\ctrTok\sigma'\rrb$ } && \ctrRule{intra--bias} \\
    \ce{ M$_{i+\frac13}$ + A{:}$\llb\ctrTok\sigma\tokSucc\tokSucc\mkrPosR\rrb$ &<=> $\llb\ctrTok\sigma\tokSucc\mkrPosL\rrb${:}B + M$_{i+\frac23}$ } && \ctrRule{pos--step} \\
    \ce{ M$_{i+\frac13}$ + A{:}$\llb\tokSucc\mkrPosR\rrb$ &<=> $\llb\mkrZero\rrb${:}B + M$_{i+\frac23}$ } && \ctrRule{pos--base} \\
    \ce{ M$_{i+\frac13}$ + A{:}$\llb\mkrZero\rrb$ &<=> $\llb\mkrNegL\tokSucc\rrb${:}B + M$_{i+\frac23}$ } && \ctrRule{neg--base} \\
    \ce{ M$_{i+\frac13}$ + A{:}$\llb\mkrNegR\tokSucc\ctrTok\sigma\rrb$ &<=> $\llb\mkrNegL\tokSucc\tokSucc\ctrTok\sigma\rrb${:}B + M$_{i+\frac23}$ } && \ctrRule{neg--step} \\
  \end{align*}\end{minipage}}}
  \caption{The abstract molecular reaction scheme implementing $\ce{ X^{(n)}_k <=> X^{(n)}_{k+1} }$ and $\ce{ X^{(n)}_n <=> X^{(n-1)}_0 }$ for the unary representation case, with $\protect\mkrAnyR$ representing any marker monomer and $\sigma$, $\sigma'$ representing arbitrary strings of $\protect\tokSucc$ monomers.}
  \label{lst:ctr-unary}
\end{listing}

The simplest implementation of counters is given by representing numbers in unary, i.e.\ a Peano axiomatic approach. For a natural integer $n\in\mathbb N$, we propose a polymeric representation consisting of $n$ repeated $\tokSucc$ monomers and represented thus,
\begin{align*}
  \llb\underbrace{\tokSucc\tokSucc\cdots\tokSucc\tokSucc}_{n}\rrb.
\end{align*}
We can then introduce a subcounter $0\le k\le n$ by incorporating a marker monomer, $\mkrPosR$, placed between the monomers and which can migrate between the $n+1$ possible positions. Moreover, we can handle non-positive values of $n$ and $k$ via the alternative marker $\mkrNegR$. Whilst not strictly necessary, we also differentiate the case $n=k=0$ with a third marker monomer, $\mkrZero$. Example polymers are given for $n=-4,0,4$:
\begin{align*}
  \ce{X^{(4)}_0} &= \llb\mkrPosR\tokSucc\tokSucc\tokSucc\tokSucc\rrb, & & & \ce{X^{(-4)}_{0}} &= \llb\tokSucc\tokSucc\tokSucc\tokSucc\mkrNegR\rrb, \\
  \ce{X^{(4)}_1} &= \llb\tokSucc\mkrPosR\tokSucc\tokSucc\tokSucc\rrb, & & & \ce{X^{(-4)}_{-1}} &= \llb\tokSucc\tokSucc\tokSucc\mkrNegR\tokSucc\rrb, \\
  \ce{X^{(4)}_2} &= \llb\tokSucc\tokSucc\mkrPosR\tokSucc\tokSucc\rrb, & \ce{X^{(0)}_0} &= \llb\mkrZero\rrb, & \ce{X^{(-4)}_{-2}} &= \llb\tokSucc\tokSucc\mkrNegR\tokSucc\tokSucc\rrb, \\
  \ce{X^{(4)}_3} &= \llb\tokSucc\tokSucc\tokSucc\mkrPosR\tokSucc\rrb, & & & \ce{X^{(-4)}_{-3}} &= \llb\tokSucc\mkrNegR\tokSucc\tokSucc\tokSucc\rrb, \\
  \ce{X^{(4)}_4} &= \llb\tokSucc\tokSucc\tokSucc\tokSucc\mkrPosR\rrb, & & & \ce{X^{(-4)}_{-4}} &= \llb\mkrNegR\tokSucc\tokSucc\tokSucc\tokSucc\rrb.
\end{align*}
Notice the mirrored interpretation of marker positions as values of $k$ for positive and negative values of $n$; this is intentional and serves to simplify the implementation of the intra-$n$ biasing reactions. It should also be noted that these polymers are achiral and so their mirror images, e.g.\ $\ce{X^{(4)}_1} = \llb\tokSucc\tokSucc\tokSucc\mkrPosL\tokSucc\rrb$, are equivalent representations of the same polymer. This property greatly simplifies the implementation of the $\ce{ X^{(n)}_n <=> X^{(n-1)}_0 }$ reactions. Altogether, this implementation can be achieved with five reactions as listed in \Cref{lst:ctr-unary}. In fact, this scheme can be implemented with just four reactions by replacing $\mkrZero$ with $\mkrPosR$ and combining reactions \ctrRule{pos-step} and \ctrRule{pos-base}, but we prefer the given implementation for its symmetry. More concretely, these reactions engender the following bi-infinite Markov chain:
{\mhchemoptions{arrow-min-length=1.5em}\begin{multline*}
  \cdots \ce{<=>} \llb\mkrPosR\tokSucc\tokSucc\rrb
  \ce{<=>>} \llb\tokSucc\mkrPosR\tokSucc\rrb \ce{<=>>} \llb\tokSucc\tokSucc\mkrPosR\rrb
  \ce{<=>} \llb\mkrPosR\tokSucc\rrb \ce{<=>>} \llb\tokSucc\mkrPosR\rrb
  \ce{<=>} \llb\mkrZero\rrb \\ \llb\mkrZero\rrb
  \ce{<=>} \llb\mkrNegL\tokSucc\rrb \ce{<<=>} \llb\tokSucc\mkrNegL\rrb
  \ce{<=>} \llb\mkrNegL\tokSucc\tokSucc\rrb \ce{<<=>} \llb\tokSucc\mkrNegL\tokSucc\rrb
  \ce{<<=>} \llb\tokSucc\tokSucc\mkrNegL\rrb \ce{<=>} \cdots.
\end{multline*}}

\subsection{Binary Representation}

{\def\AutLabel#1{{\small\textsc{#1}}}
\def\autLabel#1{\makebox[0pt][c]{\AutLabel{#1}}}

\begin{figure}
  \centering
  \begin{tikzpicture}[shorten >=1pt,node distance=2.5cm,on grid,auto]
    \node[state,initial] (succ) {\autLabel{inc}};
    \node[state,accepting] (halt) [right=of succ] {\autLabel{halt}};
    \path[->] (succ) edge [loop above] node {$1\mapsto0$; \emph{left}} ()
                     edge [bend left, above] node {$0\mapsto1$} (halt)
                     edge [bend right, below] node {$\varnothing\mapsto1$} (halt);
  \end{tikzpicture}
  \caption{The state diagram for a Turing Machine~\cite{turing-machine} implementing incrementation of a natural number in binary positional notation. The alphabet is $\{\varnothing,0,1\}$, with $\varnothing$ indicating a blank square; numbers may be provided with any number of leading 0s, and the initial position should be the least significant digit.}
  \label{fig:tm-succ}
\end{figure}
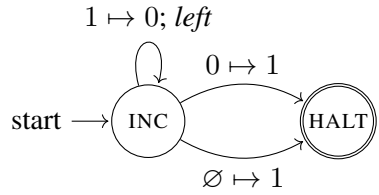

\begin{figure}
  \centering
  \begin{tikzpicture}[shorten >=1pt,node distance=2cm,on grid,auto]
     \node[state,accepting] (init) {\autLabel{init}};%
     \node[state] (carry) [right=1.5cm of init] {\autLabel{car}};%
     \node[state] (inc) [right=1.5cm of carry] {\autLabel{inc}};%
     \node[state] (end1) [above=of inc] {\autLabel{end\smash{$_1$}}};%
     \node[state] (end2) [right=of end1] {\autLabel{end\smash{$_2$}}};%
     \node[state] (dne1) [below=of inc] {\autLabel{end\smash{$_1'$}}};%
     \node[state] (dne2) [right=of dne1] {\autLabel{end\smash{$_2'$}}};%
     \node[state] (inc2) [right=of inc] {\autLabel{inc\smash{$'$}}};%
     \node[state] (inc3) [right=of inc2] {\autLabel{inc\smash{$''$}}};%
     \node[state] (carry2) [right=1.5cm of inc3] {\autLabel{car\smash{$'$}}};%
     \node[state] (carry3) [right=1.5cm of carry2] {\autLabel{car\smash{$''$}}};%
     \node[state,accepting] (term) [right=1.5cm of carry3] {\autLabel{term}};
     \node (anchor) [right=1.5cm of term] {};
    \path[->] (init) edge node {$\varnothing$} (carry)
              (carry) edge[bend left, above] node {\emph{left}} (inc)
              (inc) edge[bend left, below] node {$1 \mapsto 0$} (carry)
                    edge node [right] {$\varnothing \mapsto 1$} (end1)
                    edge node [right] {$0 \mapsto 1$} (dne1)
              (end1) edge node [above] {\emph{left}} (end2)
              (end2) edge node [right] {$\varnothing$} (inc2)
              (dne1) edge node [below] {\emph{left}} (dne2)
              (dne2) edge [bend left] node [right] {$0$} (inc2)
                     edge [bend right] node [left] {$1$} (inc2)
              (inc2) edge node [above] {\emph{right}} (inc3)
              (inc3) edge node {$1$} (carry2)
              (carry2) edge[bend left, above] node {\emph{right}} (carry3)
              (carry3) edge [bend left, below] node {$0$} (carry2)
                      edge node {$\varnothing$} (term);
  \end{tikzpicture}
\\[-3em]
  \begin{tikzpicture}[shorten >=1pt,node distance=2cm,on grid,auto]
     \node[state,accepting] (init) {\autLabel{init}};%
     \node[state] (carry) [left=1.5cm of init] {\autLabel{car}};%
     \node[state] (inc) [left=1.5cm of carry] {\autLabel{inc}};%
     \node[state] (end1) [above=of inc] {\autLabel{end\smash{$_1$}}};%
     \node[state] (end2) [left=of end1] {\autLabel{end\smash{$_2$}}};%
     \node[state] (dne1) [below=of inc] {\autLabel{end\smash{$_1'$}}};%
     \node[state] (dne2) [left=of dne1] {\autLabel{end\smash{$_2'$}}};%
     \node[state] (inc2) [left=of inc] {\autLabel{inc\smash{$'$}}};%
     \node[state] (inc3) [left=of inc2] {\autLabel{inc\smash{$''$}}};%
     \node[state] (carry2) [left=1.5cm of inc3] {\autLabel{car\smash{$'$}}};%
     \node[state] (carry3) [left=1.5cm of carry2] {\autLabel{car\smash{$''$}}};%
     \node[state,accepting] (term) [left=1.5cm of carry3] {\autLabel{term}};
     \node (anchor) [left=1.5cm of term] {};
    \path[->] (carry) edge node [above] {$\varnothing$} (init)
                      edge [bend left, below] node {$1 \mapsfrom 0$} (inc)
              (inc) edge [bend left, above] node {\emph{right}} (carry)
              (end1) edge node [left] {$\varnothing \mapsfrom 1$} (inc)
              (end2) edge node [above] {\emph{right}} (end1)
              (dne1) edge node [left] {$0 \mapsfrom 1$} (inc)
              (dne2) edge node [below] {\emph{right}} (dne1)
              (inc2) edge node [left] {$\varnothing$} (end2)
                     edge [bend left] node [left] {$0$} (dne2)
                     edge [bend right] node [right] {$1$} (dne2)
              (inc3) edge node [above] {\emph{left}} (inc2)
              (carry2) edge node [above] {$1$} (inc3)
                       edge [bend left, below] node {$0$} (carry3)
              (carry3) edge [bend left, above] node {\emph{left}} (carry2)
              (term) edge node [above] {$\varnothing$} (carry3);
  \end{tikzpicture}
  \\[1em]
  \caption{The state diagram for a Reversible Turing Machine~\cite{bennett-tm} (and its reverse) implementing incrementation (resp.\ decrementation) of a natural number in binary positional representation. The alphabet is $\{\varnothing,0,1\}$, with $\varnothing$ indicating a blank square; numbers must be provided in the described prefix-free form, and the initial position should be the (empty) square to the immediate right of the least significant digit. The box-shaped `subroutine' is used to handle the case when a new digit must be prepended; as this is a reversible branch, we must ensure that the converged state \AutLabel{inc$'$} is able to uniquely determine the branch from whence it came.}
  \label{fig:rtm-succ}
\end{figure}
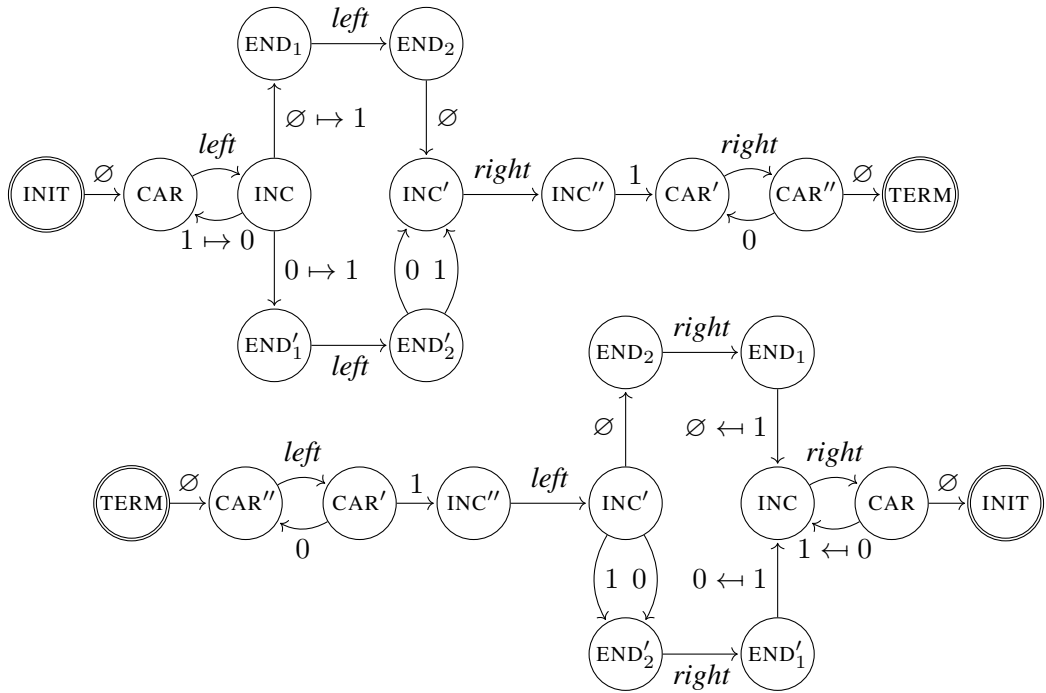}

\begin{listing}
  \centering
  \fbox{\begin{minipage}{.5\linewidth}\centering
  \def\quo#1{{\text{`}#1}}%
  \def\inc{\infix{\atom{Inc}}}%
  \def\eq#1#2{\ooalign{\hss\phantom{$#1$}\hss\cr\hss$#2$\hss}}%
  \vspace{.5em}
  \begin{align*}
    & \rlap{${[]}$}\phantom{[\quo{\eq10}~{x}~{\cdot}~{\cursivebS}]}~\inc~{[\quo{\eq01}]}{;} \\
    & {[\quo{\eq10}~{x}~{\cdot}~{\cursivebS}]}~\inc~{[\quo{\eq01}~{x}~{\cdot}~{\cursivebS}]}{;} \\
    & {[\quo{\eq01}~\phantom{x}~{\cdot}~{\cursivebS}]}~\inc~{[\quo{\eq10}~\phantom{x}~{\cdot}~{\cursivebS'}]}{:} \\
    &\qquad {\cursivebS}~\inc~{\cursivebS'}{.}
  \end{align*}
  \vspace{-.2em}
  \end{minipage}}
  \caption{A recursive $\aleph$/\texttt{alethe}~\cite{earley-aleph} program implementing incrementation of a natural number in binary positional representation. Numbers are provided in the described prefix-free form as a list of digits $\{0, 1\}$ with the least-significant bit first. The second case pattern matches on two bits in order to ensure its output pattern is distinct from that of the first case, and this is where a program will stall should a number be provided which is not in prefix-free form.}
  \label{lst:alethe-succ}
\end{listing}

\begin{listing}
  \centering
  {\longce\fbox{\begin{minipage}{0.85\textwidth}\begin{align*}
    \begin{cB}\sigma \\ \tau \\\ctrBinVar{1-1}\end{cB} &\ce{<=>>} \begin{cB*}\sigma & \ctrBinKL \\ \tau & \\\ctrBinVar{1-1}\end{cB*} && \ctrRule{$k$--init} \\
    \begin{cB*}\sigma & x & \ctrBinKL & \sigma' \\ \tau & 1 & & \tau' \\\ctrBinVar{1-1}\end{cB*} &\ce{<=>>} \begin{cB*}\sigma & \ctrBinKL & x & \sigma' \\ \tau & & 0 & \tau' \\\ctrBinVar{1-1}\end{cB*} && \ctrRule{$k$--carry} \\
    \begin{cB*}\sigma & x & 1 & \ctrBinKL & \sigma' \\\ctrBinInv{1-2} \tau & x & 0 & & \tau' \\\end{cB*} &\ce{<=>>} \begin{cB*}\sigma & x & 1 & \ctrBinKR & \sigma' \\\ctrBinInv{1-3} \tau & x & 1 & & \tau' \\\end{cB*} && \ctrRule{$k$--inc$_1$} \\
    \begin{cB*}\sigma & x & y & \ctrBinKL & \sigma' \\ \tau & 0 & 0 & & \tau' \\\ctrBinInv{1-3}\end{cB*} &\ce{<=>>} \begin{cB*}\sigma & x & y & \ctrBinKR & \sigma' \\ \tau & 0 & 1 & & \tau' \\\ctrBinInv{1-2}\end{cB*} && \ctrRule{$k$--inc$_2$} \\
    \begin{cB*}\sigma & x & y & \ctrBinKL & \sigma' \\\ctrBinVar{1-1} \tau & z & 0 & & \tau' \\\end{cB*} &\ce{<=>>} \begin{cB*}\sigma & x & y & \ctrBinKR & \sigma' \\\ctrBinVar{1-1} \tau & z & 1 & & \tau' \\\end{cB*} && \ctrRule{$k$--inc$_3$} \\
    \begin{cB*}1 & \ctrBinKL & \sigma' \\ 0 & & \tau' \\\ctrBinInv{1-1}\end{cB*} &\ce{<=>>} \begin{cB*}1 & \ctrBinKR & \sigma' \\\ctrBinInv{1-1} 1 & & \tau' \\\end{cB*} && \ctrRule{$k$--inc$_4$} \\
    \begin{cB*}\sigma & x & \ctrBinKR & 0 & \sigma' \\\ctrBinInv{1-2} \tau & x & & 0 & \tau' \\\end{cB*} &\ce{<=>>} \begin{cB*}\sigma & x & 0 & \ctrBinKR & \sigma' \\\ctrBinInv{1-3} \tau & x & 0 & & \tau' \\\end{cB*} && \ctrRule{$k$--carry$_1'$} \\
    \begin{cB*}\sigma & x & \ctrBinKR & 1 & \sigma' \\\ctrBinInv{1-2} \tau & x & & 0 & \tau' \\\end{cB*} &\ce{<=>>} \begin{cB*}\sigma & x & 1 & \ctrBinKR & \sigma' \\\ctrBinInv{1-2} \tau & x & 0 & & \tau' \\\end{cB*} && \ctrRule{$k$--carry$_2'$} \\
    \begin{cB*}\sigma & x & \ctrBinKR & z & \sigma' \\ \tau & y & & 0 & \tau' \\\end{cB*} &\ce{<=>>} \begin{cB*}\sigma & x & z & \ctrBinKR & \sigma' \\ \tau & y & 0 & & \tau' \\\end{cB*} && \ctrRule{$k$--carry$_3'$} \\
    \begin{cB*}\sigma & \ctrBinKR \\\ctrBinVar{1-1} \tau & \\\end{cB*} &\ce{<=>>} \begin{cB}\sigma \\\ctrBinVar{1-1} \tau \\\end{cB} && \ctrRule{$k$--term} \\
  \end{align*}\end{minipage}}}
  \caption{The abstract molecular reaction scheme implementing $\ce{ X^{(n)}_k <=>> X^{(n)}_{k+1} }$ for the binary representation case. To only use a single bias token's worth of free energy, one can make use of a similar scheme to that used in \Cref{eqn:dyn-coupling-scheme}, and perhaps making some of the reactions in this scheme unbiased. Alternatively one can accept the use of many tokens, which has the advantage of increasing the rate of the coupled reaction due to the effectively higher bias, but has the disadvantage of slower convergence of the sequestration klona to steady state---potentially interfering with correct operation of the coupled reaction during periods of high flux. In this scheme, the $\langle$ and $\rangle$ symbols represent klona that mark the current state and progress of the overall reaction. It is of note that, in the \ctrRule{$k$--carry} rule, there is no provision for $Q$ to be set on the bit to the left of $\begin{smallarray}{c}x\\1\end{smallarray}$, as such a state would imply we are trying to increment $k\ge n$ and is thus illegal.}
  \label{lst:ctr-bin-k}
\end{listing}

\begin{listing}
  \centering
  {\longce\fbox{\begin{minipage}{0.85\textwidth}\begin{align*}
    \ce{ M$_{i+\frac23}$ + B{:} }\begin{cB}\sigma \\ \tau \\\ctrBinInv{1-1}\end{cB} &\ce{<<=>} \begin{cB*}\sigma & \ctrBinNL \\ \tau & \\\ctrBinInv{1-1}\end{cB*}\ce{ {:}{\abmTS} } && \ctrRule{$n$--init} \\
    \ce{ {\abmTS}{:} }\begin{cB*}\sigma & 1 & \ctrBinNL & \sigma' \\\ctrBinInv{4-4} \tau & 0 & & \tau' \\\ctrBinInv{1-2}\end{cB*} &\ce{<=>} \begin{cB*}\sigma & \ctrBinNL & 0 & \sigma' \\\ctrBinInv{3-4} \tau & & 0 & \tau' \\\ctrBinInv{1-1}\end{cB*}\ce{ {:}{\abmTS} } && \ctrRule{$n$--carry}\\
    \ce{ {\abmTS}{:} }\begin{cB*}\sigma & x & 0 & \ctrBinNL & \sigma' \\\ctrBinInv{5-5} \tau & 0 & 0 & & \tau' \\\ctrBinInv{1-3}\end{cB*} &\ce{<=>} \begin{cB*}\sigma & x & \ctrBinPL & 1 & \sigma' \\\ctrBinInv{4-5} \tau & 0 & & 1 & \tau' \\\ctrBinInv{1-2}\end{cB*}\ce{ {:}{\abmTS} } && \ctrRule{$n$--inc$_1$}\\
    \ce{ {\abmTS}{:} }\begin{cB*}\ctrBinNL & \sigma' \\\ctrBinInv{2-2} & \tau' \\\end{cB*} &\smash{\ceArrAdd{<=>}{$\mathbb B_2$}} \begin{cB*}\ctrBinPL & 1 & \sigma' \\\ctrBinInv{2-3} & 1 & \tau' \\\end{cB*}\ce{ {:}{\abmTS} } && \ctrRule{$n$--inc$_2$}\\
    \ce{ {\abmTS}{:} }\begin{cB*}\sigma & x & \ctrBinPL & y & \sigma' \\\ctrBinVar{5-5} \tau & 0 & & 0 & \tau' \\\ctrBinInv{1-2}\end{cB*} &\ce{<=>} \begin{cB*}\sigma & \ctrBinPL & x & y & \sigma' \\\ctrBinVar{5-5} \tau & & 0 & 0 & \tau' \\\ctrBinInv{1-1}\end{cB*}\ce{ {:}{\abmTS} } && \ctrRule{$n$--prefix$_1$}\\
    \ce{ {\abmTS}{:} }\begin{cB*}\ctrBinPL & \sigma' \\\ctrBinVar{2-2} & \tau' \\\end{cB*} &\ce{<=>} \begin{cB*}\ctrBinPR & \sigma' \\\ctrBinVar{2-2} & \tau' \\\end{cB*}\ce{ {:}{\abmTS} } && \ctrRule{$n$--prefix$_2$}\\
    \ce{ {\abmTS}{:} }\begin{cB*}\sigma & \ctrBinPR & x & \sigma' \\\ctrBinInv{1-1}\ctrBinVar{4-4} \tau & & 0 & \tau' \\\end{cB*} &\ce{<=>} \begin{cB*}\sigma & x & \ctrBinPR & \sigma' \\\ctrBinInv{1-2}\ctrBinVar{4-4} \tau & x & & \tau' \\\end{cB*}\ce{ {:}{\abmTS} } && \ctrRule{$n$--prefix$_3$}\\
    \ce{ {\abmTS}{:} }\begin{cB*}\sigma & \ctrBinPR & 1 & \sigma' \\\ctrBinInv{1-1}\ctrBinVar{4-4} \tau & & 1 & \tau' \\\end{cB*} &\ce{<=>} \begin{cB*}\sigma & 1 & \ctrBinNR & \sigma' \\\ctrBinInv{1-2}\ctrBinVar{4-4} \tau & 1 & & \tau' \\\end{cB*}\ce{ {:}{\abmTS} } && \ctrRule{$n$--inc$'$}\\
    \ce{ {\abmTS}{:} }\begin{cB*}\sigma & \ctrBinNR & 0 & \sigma' \\\ctrBinInv{1-1}\ctrBinInv{3-4} \tau & & 0 & \tau' \\\end{cB*} &\ce{<=>} \begin{cB*}\sigma & 0 & \ctrBinNR & \sigma' \\\ctrBinInv{1-2}\ctrBinInv{4-4} \tau & 0 & & \tau' \\\end{cB*}\ce{ {:}{\abmTS} } && \ctrRule{$n$--carry$'$}\\
    \ce{ {\abmTS}{:} }\begin{cB*}\sigma & \ctrBinNR \\\ctrBinInv{1-1} \tau & \\\end{cB*} &\ce{<=>>} \begin{cB}\sigma \\\ctrBinInv{1-1} \tau \\\end{cB}\ce{  {:}A + M$_{i+\frac13}$ } && \ctrRule{$n$--term} \\
  \end{align*}\end{minipage}}}
  \caption{The abstract molecular reaction scheme implementing $\ce{ X^{(n)}_0 <=> X^{(n+1)}_{n+1} }$ for the binary representation case. This overall reaction is unbiased, which raises the concern that the sequestration klona could spend an undesirable amount of time in the intermediate states above. To limit this possibility, the intermediate states should be made more unfavourable and this is achieved by the complementary biases in the rules \ctrRule{$n$--init} and \ctrRule{$n$--term}. These biases need not be provided by the bias tokens, and in fact that is unideal as the bias is vanishingly small; instead it can be implemented by introducing an enthalpy change, such that there is a chance of thermal activation of the reaction, whilst the klona still prefer to exist in the non-intermediate states. As in the scheme for the intra-$n$ reaction, klona are introduced to mark the state and progress of the overall reaction; in this case, $\{$, $\}$, $($, and $)$, with the $\{$ and $\}$ klona corresponding primarily to incrementation of $n$ and the $($ and $)$ klona to copying of the prefix of $n$ into $k$.}
  \label{lst:ctr-bin-n}
\end{listing}

Whilst particularly simple, the unary representation is very space inefficient; worse, incrementing and decrementing $n$, for the purpose of modulating access to a sparse resource/disequilibrium reaction, will itself require interaction with a resource pool in order to acquire and release $\circ$ monomers. Fortunately an exponentially more compact representation exists in the form of positional notations such as binary, with spatial complexity logarithmic in $|n|$ and hence needing far fewer interactions with its respective monomer pool.

Unlike with the unary case, we cannot exploit the structure of the representation of $n$ to represent $k$. Instead we shall need to store both explicitly. We will also require the ability to recognise the states $k=0$ and $k=n$, both for determining whether the coupled reaction $\ce{A <=> B}$ should proceed and for ensuring the reactions that generate the intra-$n$ distribution do not produce illegal values of $k$. These two states correspond respectively to all the digits of $k$ being $0$ or being identical to those of $n$. Since $|k|<|n|$, its minimal representation will be no larger than that of $n$ and hence a convenient structure for encoding the pair $(n,k)$ with the ability to easily check the condition $n=k$ is given by a `zipped' double-stranded polymer which we denote by $\llb\begin{smallarray}{c}n\\k\end{smallarray}\rrb$; for example, in base 2 the pair $(11,6)$ would be given by
\[ \begin{cB}1 & 0 & 1 & 1 \\ 0 & 1 & 1 & 0 \\\end{cB}. \]
In contrast to the unary case, this representation \emph{does} have an intrinsic polarity, with the leftmost digits being the most significant. To ensure logical reversibility of the forthcoming reaction scheme, the polymer representation must be unique for each $(n,k)$ pair, but positional representations admit a countably infinite equivalence class for each integer (e.g.\ 3 has base-10 representations $\{3,03,003,\ldots\}$). We resolve this ambiguity by trimming all leading zeros of $n$ and trimming $k$ to the same length. This has the consequence that $(0,0)$ has the `empty' representation $\llb\rrb$; it is certainly possible to introduce a special case for $(0,0)$ of $\llb\begin{smallarray}{c}0\\0\end{smallarray}\rrb$ for {\ae}sthetic reasons, but this would significantly complicate the reaction scheme for no other benefit. 

Thus far, we have only considered the representation for non-negative $k,n\in\mathbb N$; there are a number of approaches one could take to extend the representation, from simply incorporating a \emph{sign} trit, $\{-,0,+\}$, similar to the approach for the unary case, to a two's complement approach. The two's complement approach is particularly attractive as it would require only minimal adjustments to the non-negative reaction scheme. The two's complement representation of negative numbers is realised by taking the equivalence class for the integer, e.g. $\{11,011,0011,\ldots\}$ for 3, and complementing each bit to obtain, e.g., $\{00,100,1100,\ldots\}$. That is, each negative integer has an infinite prefix of 1s instead of 0s as for the non-negative case. This approach is used in most conventional CPUs because no special-casing for negative numbers is necessary: adding a negative number $m$, with two's complement $m'$, to a positive number $n$ is equivalent to adding the positive numbers $m'$ and $n$. As a simple justification for this, consider decrementing $1000\cdots000$. We will need to `borrow' a 1 from each digit to the left, eventually obtaining $111\cdots111$. In the limit of infinitely many 0s, this results in infinitely many 1s. For brevity we shall restrict our attention to the non-negative case, but extending to the full domain is a straightforward---if tedious---exercise.

Whilst recognising the cases $k=0$ and $k=n$ is not difficult in our representation, comparing $\sim\log n$ digits each time we wish to execute the coupled reaction is not ideal. By a slight increase in the complexity of the $\ce{ X^{(n)}_k <=>> X^{(n)}_{k+1} }$ and $\ce{ X^{(n)}_n <=> X^{(n-1)}_0 }$ algorithms, we can render these checks trivial. In particular, we augment the polymer with two bits of state for each monomer pair which we call $Q$, for e$Q$uality, and $Z$, for $Z$ero. As biochemical precedent, compare with phosphorylation sites. The function of these states is best understood by example; for $n=11$, the $Q$-$Z$ states for each of $k=0,1,8,11$ are given by
\begin{align*}
  \begin{cB}1 & 0 & 1 & 1 \\ 0 & 0 & 0 & 0 \\\ctrBinInv{1-4}\end{cB}, &&
  \begin{cB}1 & 0 & 1 & 1 \\ 0 & 0 & 0 & 1 \\\ctrBinInv{1-3}\end{cB}, &&
  \begin{cB}1 & 0 & 1 & 1 \\\ctrBinInv{1-2} 1 & 0 & 0 & 0 \\\end{cB}, &&
  \begin{cB}1 & 0 & 1 & 1 \\\ctrBinInv{1-4} 1 & 0 & 1 & 1 \\\end{cB}.
\end{align*}
Namely, a prefix of zeroes is marked by the $Z$ state (bottom lines) and a prefix matching $n$ is marked by the $Q$ state (middle lines), and these are easily achieved by implementing two local invariants for each digit pair $\begin{smallarray}{c}x\\y\end{smallarray}$:
\begin{enumerate}
  \item[$(Q)$] The $Q$ state is \emph{on} if and only if $y=x$ and the $Q$ state of the digit pair to its left (if it exists) is also \emph{on}.
  \item[$(Z)$] The $Z$ state is \emph{on} if and only if $y=0$ and the $Q$ state of the digit pair to its left (if it exists) is also \emph{on}.
\end{enumerate}
It can be seen that these two states are orthogonal as the leading bit of $n$ must be 1, with the possible exception of $(0,0)$ where it makes sense to define $Q$ and $Z$ to be both \emph{on} despite the lack of lack of monomers on which to mark said states.

Before implementing a reaction scheme, we first require an algorithmic understanding of reversibly incrementing/decrementing integers in positional notation. As it is possible (and indeed trivial) to do so for integers in unary notation as demonstrated by \Cref{lst:ctr-unary} we should expect it to also be possible for positional notation, and indeed it is. The irreversible algorithm for incrementing a binary number is very simple (\Cref{fig:tm-succ}): if there are any trailing 1 bits then we carry the 1 (flipping the intermediate 1 bits to 0) until we reach the first 0, which we flip to 1. If we reach the end of the number, we prepend a new 1 bit. Making this algorithm reversible is conceptually simple, but expressing it in the form of a Reversible Turing Machine (RTM, as introduced by \textcite{bennett-tm}) is somewhat involved. An example implementation is provided in \Cref{fig:rtm-succ}, and can be broken down into three stages: (1) carrying; (2) incrementing the least significant 0 bit, or prepending a fresh 1 bit, followed by the logic to reversibly join these two branches; (3) `reverse' carrying, in which we return to the starting position and use the fact that the suffix we are processing takes the form $100\cdots000$ to ensure we can logically reverse this stage. It can be readily checked that this scheme as presented is reversible, and indeed we provide the mechanical reversal of the RTM which may be seen to implement the operation of reversibly decrementing a positive natural number. To demonstrate that the algorithm is in fact conceptually simple, an equivalent recursive program written in a higher level reversible language is shown in \Cref{lst:alethe-succ}.

Finally, the reaction schemes for $\ce{ X^{(n)}_k <=>> X^{(n)}_{k+1} }$ and $\ce{ X^{(n)}_n <=> X^{(n-1)}_0 }$ are made manifest in \Cref{lst:ctr-bin-k,lst:ctr-bin-n} respectively. In addition to implementing the reversible incrementation procedure in an abstract molecular sense, these schemes must also incorporate the relevant logic for ensuring $k$ is appropriately bounded and that the $Q$ and $Z$ invariants are preserved (although these may be transiently broken within the intermediate states). Furthermore the logic for the second reaction (which we choose to implement in reverse, i.e.\ $\ce{ X^{(n)}_0 <=> X^{(n+1)}_{n+1} }$, per the freedom reversibility affords us) must reversibly copy the value of $n$ into $k$, and thus the second scheme must read the entire length of the polymer. As for the case of $n\equiv2^p-1$ in $n\mapsto n+1$ (resp. $n+1\mapsto n$), the scheme recruits (resp.\ releases) a pair of binary monomers ($\mathbb B_2$) which will be provided by some resource pool mediated by its own sequestration klona. To prevent the molecular algorithms running multiple times at once, the intermediate states are distinguished via an additional state in the form $\llb\,\cdot\,\ooalign{\hss\hspace{-0.2ex}$\bullet$\hss\cr$\rrbracket$}$.

\printbibliography

\end{document}